\documentclass[10pt,a4paper]{article} 
\usepackage{amsmath}
\usepackage{cite}
\usepackage{booktabs}
\usepackage{array}
\usepackage{graphicx}
\usepackage{url}

\begin{document}

\title{Evaluating Model Checking Approaches to Verify Stability of Control Systems in Simulink}

\author{Dejanira Araiza-Illan and Kerstin Eder\\
Department of Computer Science\\ University of Bristol, Bristol, UK \\
Email:\texttt{\{Dejanira.Araizaillan, Kerstin.Eder\}@bristol.ac.uk}}
\date{}

\maketitle

\begin{abstract}
This paper examines the verification of stability, a control requirement, over discrete control systems represented as Simulink diagrams, using different model checking approaches and tools.  
Model checking comprises the (exhaustive) exploration of a model of a system, to determine if a requirement is satisfied. 
If that is not the case, examples of the requirement's violation within the system's model are provided, as witnesses. 
These examples are potentially complementary to previous work on automatic theorem proving, when a system is not proven to be stable, but no proof of instability can be provided. 

We experimentally evaluated the suitability of four model checking approaches to verify stability on a set of benchmarks including linear and nonlinear, controlled and uncontrolled, discrete systems, via Lyapunov's second method or Lyapunov's direct method. 
Our study included symbolic, bounded, statistical and hybrid model checking, through the open-source tools NuSMV, UCLID, S-TaLiRo and SpaceEx, respectively. 
Our experiments and results provide an insight on the strengths and limitations of these model checking approaches for the verification of control requirements for discrete systems at Simulink level. 
We found that statistical model checking with S-TaLiRo is the most suitable option to complement our previous work on automatic theorem proving. 
\end{abstract}


\section{INTRODUCTION}
The {\em verification} of control systems is a timely need, especially for complex autonomous systems interacting with people, such as autonomous cars, or service robots. 
Verification processes allow gaining confidence and gathering evidence that the designed systems work according to functional requirements~\cite{IEEEstandard} or that they are {\em dependable}. 
Requirements are grouped into safety (``nothing bad ever happens'') or liveness (``something good eventually happens''). 

Control systems need to be verified at all phases in the design process from conceptual ideas to code implementation. 
Along these phases, different requirements need to be considered for verification.
Whereas at the initial and mathematical abstract design phases, theoretical control and functional requirements such as stability or robustness need to be verified, further requirements such as absence of runtime errors (arrays out of bounds, variable overflows) and floating-point issues arise at code implementation, the lowest levels. 
It is important to: \textit{(a)} verify requirements early, at the right level of abstraction and optimal design-to-implementation phase; and \textit{(b)} follow a ``design and implementation for verification'' philosophy, to facilitate the use of verification techniques and tools. 
As control systems differ according to their target systems (e.g., linear, nonlinear, stochastic, noisy, partially observable), there is a growing need to understand what existing verification techniques can deliver for the verification of these systems.

In our work, we have focused on the verification of discrete control systems modelled at Simulink level from difference equations, assumed to lay in between a pen-and-paper theoretical design phase, and a code implementation derived automatically through MATLAB or developed by hand. 
We look into the use and combination of well established verification techniques (testing, theorem proving, model checking), control systems theory, and accomplished and supported tools, targeting these discrete systems.

Previously, we investigated the combination of two verification techniques, {\em theorem proving} and {\em numerical tests}~\cite{Araiza:2014,Araiza:2015}, to verify control requirements such as stability in the most automatic manner possible, and ``for all possible variable values and initial conditions''.
Theorem proving allows computing a mathematical proof of the requirements in a symbolic and static manner, i.e., without running simulations. 
However, if the theorem proving fails to compute a proof, no additional information is provided, making debugging difficult. 
In contrast, other verification techniques such as {\em model checking} and {\em testing} provide witnesses or evidence of a violation of the requirements. 
Nevertheless, both are computationally expensive, as model checking exhaustively explores a model, and testing might require a large number of simulations to achieve an acceptable level of coverage of the system.

Transparency in the systems to observe relevant variables and parameters, and the formulation of the requirement at the right abstraction level (e.g., referring to elements in a Simulink model) and in a quantifiable manner, if possible, are paramount in ``design for verification''. 
Hence, control requirements in natural language, like ``stability'', require translation into a metric in terms of a system's variables and parameters. 
The translation process requires specialist knowledge, such as control systems theory, and a degree of control over the implementation of a Simulink model. 
In our previous work~\cite{Araiza:2014,Araiza:2015}, we made use of assertions in the form of Simulink blocks, to express the requirements to verify at the Simulink diagram level. 
We incorporated Lyapunov functions into the Simulink diagrams, to assess stability. 
Although Lyapunov functions can be computed through well established procedures, more than one might be suitable for the same system. 
Automated verification procedures, such as model checking or automatic theorem proving, would help to establish, according to the proposed Lyapunov functions, if a system is stable or that no stability guarantees can be provided.
The latter can be caused by a proposed function that does not behave as a Lyapunov function, or if the system is indeed unstable. 
Translating the system into a suitable model for model checking is a challenging task, as some model checking tools only function with finite-state transition or hybrid models. 

In this paper we experimentally investigated the feasibility to verify stability for discrete systems using model checking.
Model checking has diversified to handle systems with many states and continuous components (e.g., hybrid systems). We set out to compare four different model checking approaches in a systematic manner. 
Additionally, we sought to find out whether model checking can complement previously proposed theorem proving methodologies, such as the ones proposed in~\cite{Araiza:2014,Araiza:2015}, by providing evidence of requirement satisfaction or violation in the form of witnesses.  

Our study included state-of-the-art symbolic~\cite{ClarkeMC}, bounded (BMC)~\cite{ClarkeMC}, statistical (sampling-based)~\cite{Zuliani} and hybrid~\cite{Alur} model checking approaches, through representative tools (model checkers): NuSMV\footnote{http://nusmv.fbk.eu/}, UCLID\footnote{http://uclid.eecs.berkeley.edu/}, S-TaLiRo\footnote{https://sites.google.com/a/asu.edu/s-taliro/s-taliro}, and SpaceEx\footnote{http://spaceex.imag.fr/}, respectively. 
These tools were chosen due to their announced compatibility with Simulink, where applicable; also, they are well maintained and user friendly. 
We determined the advantages and limitations of each one of these model checking approaches, with respect to the verification of stability requirements for linear scalar, linear multi-variable and nonlinear multi-variable discrete systems, all modelled as Simulink diagrams with basic blocks. 
Applied quantitative and qualitative performance criteria comprised: correctness of the translation semantics from Simulink to the input language of the model checker, time to pre-process a model into a suitable representation (if needed), time to check a requirement, and amount of additional user analysis to specify checking parameters such as simulation time or loop iterations. 

As in our previous work, we incorporated Lyapunov functions into the Simulink diagrams, formulating the Temporal Logic properties~\cite{ClarkeMC} to verify through model checking in terms of Lyapunov functions and their characteristics. 
For the linear systems, we employed Lyapunov's second method to determine stability, as in~\cite{Araiza:2015}; for the nonlinear system, we employed Lyapunov's direct method. 
We encoded suitable models for verification into the input languages of the model checkers, based on existent translation semantics.
In the models, we attempted to balance the number of states in the models on one hand, to avoid a state-space explosion, and expressing the continuity of the state-space of the discrete systems on the other.

The paper proceeds with an overview of related work on verification of Simulink diagrams, and verification of the stability control requirement in Section~\ref{sc:relatedwork}. 
We then present different case studies used as benchmarks (Section~\ref{sc:examples}), followed by a brief introduction of the main features of the  model checking tools (Section~\ref{sc:mctools}) that were employed to verify stability in each case study. 
Section~\ref{sc:experiments} presents the comparative results of the verification experiments. 
Section~\ref{sc:conclusion} concludes the paper and gives and outlook towards future work.

\section{RELATED WORK}\label{sc:relatedwork}
Verification techniques include testing, model checking and theorem proving. 
In practice, combinations of these techniques are used to verify complex real-life systems, departing from the ``one technique fits all'' paradigm. 
The presence of signals and parameters theoretically in the domain of the real numbers, corresponding to a continuous real world, leads to state-space explosion problems in the computational mechanisms of some of these techniques.

In testing, inputs are applied to a system to stimulate actions and reactions, and outputs are observed to determine if the requirements are satisfied. 
The selection of inputs (test cases) needs to thoroughly explore the system's state space, whilst targeting its interesting regions (i.e., ``covering'' the system). 
Simulink is an ideal tool for testing models of control systems in simulation. Test generation systematically samples the state space of variables and parameters, e.g., through automated search~\cite{Zhan}. 

Theorem proving or deductive verification~\cite{ClarkeMC} is a static verification technique that involves finding a mathematical proof of a requirement, through the application of axioms, lemmas and inference rules. 
A proof can be computed automatically via Satisfiability Modulo Theory (SMT) solvers or Satisfiability (SAT) solvers, or interactively (with user guidance), which requires a great degree of domain knowledge and expertise. 
A description of the system and requirements in Propositional, First-Order or Higher-Order Logic is required, along with any other relevant mathematical theory (e.g., sets, linear algebra). 
These definitions and additional information are normally encoded by hand into ``theories'', as required by case studies; they can be reused once embedded in the theorem proving tools.  
Theorem proving has been employed to verify functional equivalence between Simulink diagrams and auto-generated code (e.g.,~\cite{Circus}), for data type checks (e.g.,~\cite{Simcheck}) and to verify high-level requirements including stability (e.g.,~\cite{Zou2015}).

Model checking is the exhaustive traversal of a finite-state model of a system (i.e., all the states and state transitions in the model are explored) to check for requirements defined as properties in a variety of different Temporal Logics~\cite{ClarkeMC}. 
Hence, most model checking variants require a discrete or hybrid model that is decidable. 
If a property is found to be false, a counterexample is returned, comprising a sequence of states or a trace (according to the valid state transitions in the model).
Computing a decidable model and reducing the state-space to avoid state-space explosion issues, have motivated the shift from explicit-state model checking, i.e., enumerating and traversing all possible states, to symbolic (grouping states in compacted Binary Decision Diagrams or BDDs), bounded (exploring up to $k$ transitions in the model)~\cite{ClarkeMC}, and statistical (sampling the model's state space)~\cite{Zuliani} model checking.

Probabilistic model checking tools~\cite{Beer} suit stochastic models such as Discrete Time Markov Chains.
Specialist hybrid model checking tools -- for hybrid models comprising both discrete and continuous transitions, such as switched systems -- make use of geometrical methods to approximate the explored state space of the continuous transitions~\cite{Chutinan}.  
Hybrid model checkers (and other verification techniques such as theorem proving) commonly restrict the continuous components to ordinary differential equations (ODE) with linear or affine forms. 
Reduction of the models can be achieved by systematic abstractions (e.g., bisimulations), or symmetry reduction techniques.

The absence of runtime errors (or low-level requirements) such as overflows or arrays out of bounds for fixed data widths, was verified in~\cite{Herber2013} using model checking for Simulink diagrams. Other tools, such as Mathwork's Polyspace, translate the Simulink diagrams into code before checking for runtime errors. 
Higher-level requirements in terms of safety and liveness have been verified directly in the Simulink models (e.g., the Prover Plug-In{\textregistered}  or CheckMate for hybrid systems~\cite{Chutinan}), after translating the models (or parts of them) into the language of a specific model checker~\cite{Ezekiel,Beer}, or after translating the Simulink diagrams into  code~\cite{Barnat,Wang2014}. 
Since model checking is based on the exploration of finite-state decidable models, which implies discretization and abstraction processes over the original systems, formalized ``translation'' processes are highly desirable. 
We explored available translation semantics from Simulink to NuSMV~\cite{Meenakshi} and to UCLID~\cite{Herber2013}.

The computation of decidable models goes in hand with developing sound automated translation procedures. 
This leads to further considerations on the pros and cons of translating and verifying the Simulink models as code, potentially with runtime issues having been introduced in the process, versus considering them as mathematical control system models and to verify the absence of fundamental design flaws before runtime issues are being introduced on the way to code generation. 

Control systems requirements such as stability via Lyapunov methods have been verified mostly through theorem proving by directly posing the problems in mathematical terms (e.g., inequalities of region intersections as in~\cite{Jackson}) for continuous systems, or over controllers implemented in code~\cite{Wang2014}, closer to discrete systems. 

From a theoretical control systems perspective, model checking has been applied, via Lyapunov methods, to verify stability for particular types of continuous~\cite{Zoucav} or hybrid systems~\cite{Kapinski}. 
For practitioners, however, it is important to understand whether any model checking approach would be suitable also for more generic discrete systems, linear and nonlinear. Our paper aims to provide an insight into this.

\section{SYSTEM EXAMPLES}\label{sc:examples}
In this paper we verified stability control requirements over Simulink diagrams through model checking, to evaluate how model checking variants compare and if they could be used to complement theorem proving by providing witnesses when no proof can be found.
We chose textbook case studies from control systems theory: linear scalar, linear multi-variable, and non linear multi-variable discrete systems. 
The stability requirement was parametrized in terms of the Simulink diagrams' components through the application of Lyapunov theory, as proposed previously in~\cite{Araiza:2015}, and summarized next. 

\subsection{Lyapunov's Second Method}
Linear systems have a single equilibrium point, nonlinear systems have multiple equilibrium points.  
An equilibrium point is {\em stable} if the system's state trajectories, starting from any initial point close to the equilibrium point, remain close to it. 
An equilibrium point is {\em asymptotically stable} if it is stable and the trajectories move towards the equilibrium point as the time $t \rightarrow \infty$. 

A Lyapunov function, $V(\mathbf{x}(k))$  for a discrete system (with variables $\mathbf{x}(k)$) is a function such that:
\begin{itemize}
\item $V(\mathbf{x}(k))>0, \forall \mathbf{x}(k) \neq \mathbf{0}$, and $V(\mathbf{x}(k))=0$ if $\mathbf{x}(k)=\mathbf{0}$ (at the equilibrium point).
\item $V(\mathbf{x}(k))-V(\mathbf{x}(k-1))< 0, \forall \mathbf{x}(k) \neq \mathbf{0}$, and $V(\mathbf{x}(k))-V(\mathbf{x}(k-1))= 0$ if  $\mathbf{x}(k)=\mathbf{0}$. 
\end{itemize}

A discrete system is asymptotically stable at the equilibrium point if and only if there exists a Lyapunov function. 
For linear and hybrid systems, a candidate Lyapunov function, with $\mathbf{P}$ a positive definite matrix, is
\begin{equation}\label{lyapunovfunction}
V(\mathbf{x}(k))=\mathbf{x}(k)^{\mathrm{T}}\mathbf{Px}(k). 
\end{equation}

This function can be computed from solving a relevant Lyapunov's equation or a set of equations (for hybrid systems). 
For a nonlinear system, Lyapunov's second method can be applied after linearizing around each one of the equilibrium points, for all the resulting linear systems. 
Alternatively, a specific Lyapunov function can be proposed (Lyapunov's direct method). 

Although we can compute single Lyapunov functions given established procedures, we can propose other Lyapunov functions that might be compatible with the system. 
Furthermore, the translation of the system for which we designed the Lyapunov functions, from pen-and-paper into a Simulink diagram (or code), might be incorrect. 
Thus, automated procedures to verify stability are greatly desirable, to help ensure that the designs satisfy their control requirements.

\subsection{Linear Time Invariant Discrete Systems}

Three discrete systems were chosen: two simple uncontrolled loops, and a controlled system. We proposed suitable Lyapunov functions for each system, to facilitate the verification of stability. 

\begin{figure}[t]
\centering
\includegraphics[width=0.35\columnwidth]{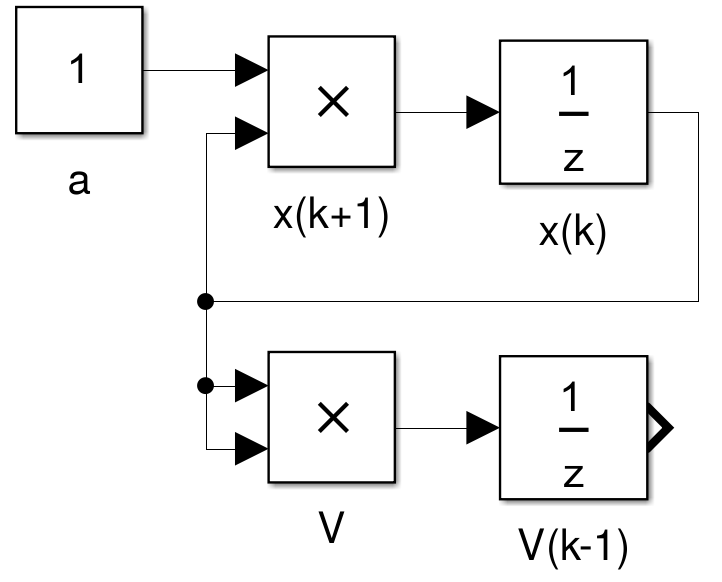} 
\caption{Linear single-variable system with Lyapunov function}
\label{system1}
\end{figure}

\paragraph{Multiplication loop} Inspired by the example in~\cite{Herber2013}, shown in Fig.~\ref{system1} and defined as
\begin{equation}
x(k+1)=ax(k).
\end{equation}
A Lyapunov function for this system is
\begin{equation}
V(x) = x^2.
\end{equation}

\begin{figure}[t]
\centering
\includegraphics[width=0.35\columnwidth]{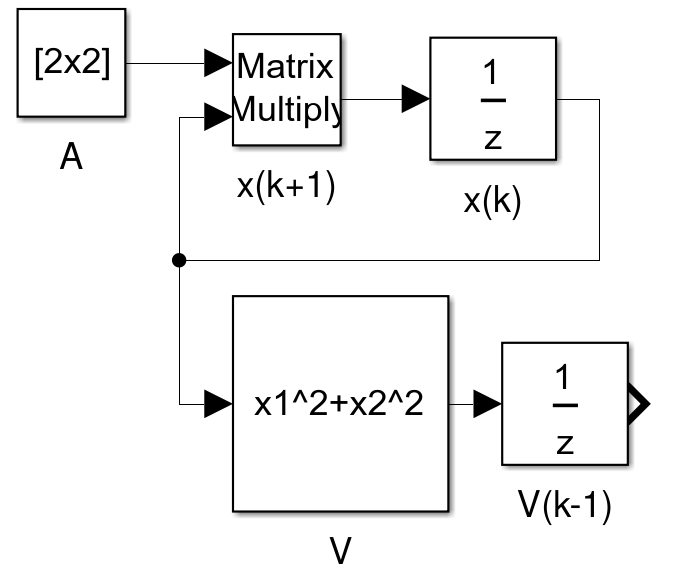} 
\caption{Linear multi-variable system with Lyapunov function}
\label{system2}
\end{figure}

\paragraph{Multi-variable loop} Example from~\cite{Araiza:2014}, shown in Fig.~\ref{system2}, and defined as
\begin{equation}
\mathbf{x}(k+1)=\mathbf{A}\mathbf{x}(k). 
\end{equation}
A Lyapunov function was proposed according to~(\ref{lyapunovfunction}), with $\mathbf{P}=\mathbf{I}$.

\begin{figure}[t]
\centering
\includegraphics[width=0.55\columnwidth]{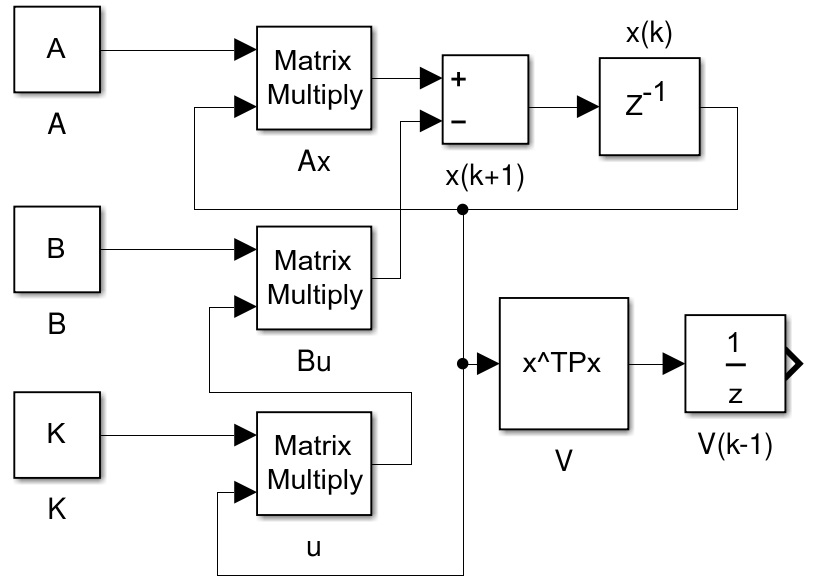} 
\caption{Controlled linear multi-variable system with Lyapunov function}
\label{system3}
\end{figure}

\paragraph{Multi-variable controlled system} Example from~\cite{Araiza:2015}, shown in Fig.~\ref{system3}. In state-space equation form, defined as
\begin{equation}
\mathbf{x}(k+1)=\mathbf{A}\mathbf{x}(k)+\mathbf{Bu}(k) \label{system:loop},
\end{equation}
with matrices
\setlength{\arraycolsep}{0.4em}
\begin{eqnarray}
\mathbf{A}&=&\left[ \begin{array}{cc}
1.5&0.5\\ 0.5&1 \end{array}\right], \ \ \mathbf{B}=\left[ \begin{array}{r}
2\\ 0 \end{array} \right], \nonumber
\end{eqnarray}
and a feedback controller for stability,
\begin{equation}\label{controller}
\mathbf{u}(k)=-\mathbf{K} \mathbf{x}(k), \ \  \mathbf{K}=\left[\begin{array}{cc} 1.15&0.57\end{array}\right],
\end{equation}
by pole placement with desired poles $[0.8,0.3;0 -0.6]$. A Lyapunov function was proposed according to~(\ref{lyapunovfunction}), computed from a Lyapunov's discrete equation, 
\begin{equation}\label{disclyap}
\mathbf{(A-BK)}^{\mathrm{T}}\mathbf{P(A-BK)}-\mathbf{P}=-\mathbf{I},
\end{equation}
\begin{equation}
 \mathbf{P}= \left[ \begin{array}{cc}
2.26&1.50\\ 1.50&4.06 \end{array}\right].\nonumber
 \end{equation} 

These systems represent infinite loops computationally. If the systems are stable, fixpoints can be derived for the system variables, in the equilibrium points. If the systems are unstable, the system variables will grow without bounds, a challenge for automated verification tools such as model checking, in terms of state-space explosion, data representation (overflows) and decidability (procedures to exhaustively explore a system's model might not terminate). 

\subsection{Nonlinear Discrete System}
\begin{figure}[t]
\centering
\includegraphics[width=0.85\columnwidth]{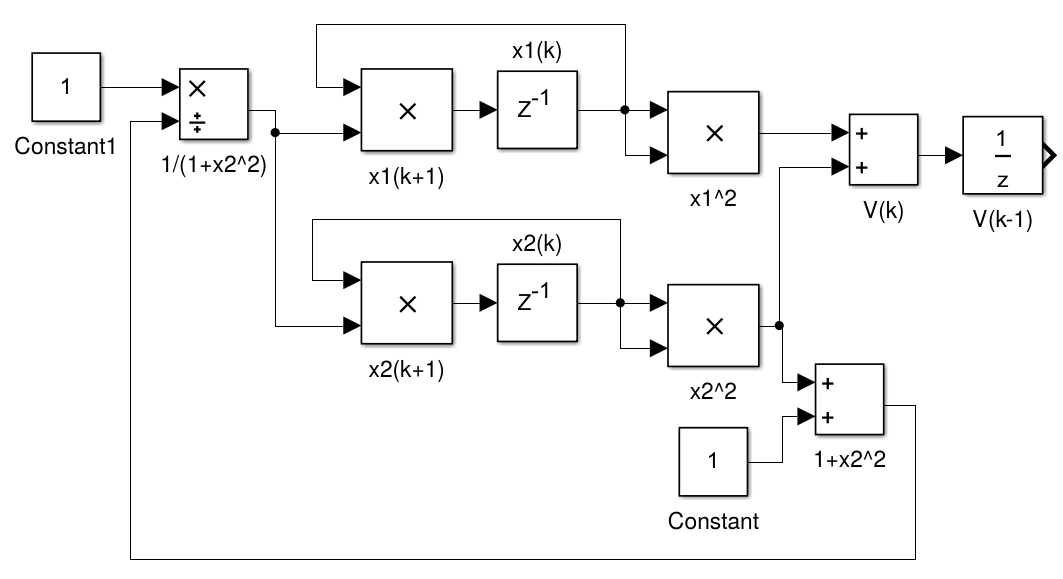} 
\caption{Nonlinear multi-variable system with Lyapunov function}
\label{system4}
\end{figure}

The selected nonlinear discrete system is shown in Fig.~\ref{system4}, defined as
\begin{eqnarray}
x_1(k+1) &=& \frac{x_2(k)}{1+x_2^2(k)} \nonumber \\ 
x_2(k+1) &=& \frac{x_1(k)}{1+x_2^2(k)}.
\end{eqnarray}
A Lyapunov function was proposed,
\begin{equation}
V(\mathbf{x}(k))=x_1^2(k) + x_2^2(k),
\end{equation}
with the difference 
\begin{equation}
V(\mathbf{x}(k))-V(\mathbf{x}(k-1)) = V(\mathbf{x}(k))\left[\frac{1}{[1+x_2^2(k)]^2}-1 \right] \leq 0. \nonumber
\end{equation}

Computationally, nonlinear system loops employ arithmetic operations that might lead to errors, such as divisions by zero, in automated verification processes. This is added to the aforementioned state-space explosion issues.

\section{SELECTED MODEL CHECKING TOOLS}\label{sc:mctools}
Specific model checking tools, corresponding to different model checking approaches, were chosen following the criteria of: 
\textit{(a)} widespread usage within the verification community;
\textit{(b)} good support, to ensure fully functioning tools and, thus, high productivity; 
\textit{(c)} user friendliness, providing guides and examples; and 
\textit{(d)} previous application for the verification of Simulink diagrams. 

\subsection{NuSMV}
NuSMV, a symbolic model checker, was originally developed for the verification of hardware designs. 
A finite-state machine (FSM) describes the model, in terms of states and their transitions. 
The transition model is transformed into a Boolean function, which is encoded into a BDD, a data structure developed for compressed Boolean function representations~\cite{ClarkeMC}. 
Consequently, the encoded BDD structures are not efficient if the transitions are dictated by complex arithmetic operations, such as a series of multiplications in a control system loop. 

The requirements to verify can be encoded into Linear Temporal Logic (LTL) or Computation Tree Logic (CTL)~\cite{ClarkeMC}. 
Also, NuSMV can perform BMC, by specifying a maximum number of transitions to explore in the model.

NuSMV's syntax includes signed and unsigned bit-vectors, and matrix data types (arrays of arrays of bit-vectors). 
Translation semantics from Simulink to NuSMV have been proposed in~\cite{Meenakshi}.

\subsection{UCLID}

UCLID was also originally developed for the verification of hardware designs. 
Its syntax is similar to NuSMV's, although it does not include matrix data types and the arithmetic operations have a more limited functionality (e.g., the division operation only allows integers to the power of 2 as denominators). 
The models have a FSM form, but they are not encoded as BDDs. 

UCLID performs BMC, by ``simulating'' the FSM for a specified number of transitions, and checking a specified logical-mathematical expression at each step or once all steps have finished. 

Translation semantics from Simulink to UCLID have been proposed in~\cite{Herber2013}.

\subsection{S-Taliro}
S-TaLiRo is categorized within statistical model checking, since sample traces or state sequences are extracted from the model to determine if a Metric Temporal Logic (MTL) property is true or false in all these samples~\cite{Zuliani}. 
Its goal is finding (sequences of) inputs in a Simulink diagram that satisfy (or falsify) a requirement in terms of outputs in the diagram, or the verification of the requirement according to a range of initial system parameters (e.g., initial states $\mathbf{x}(0)$). 
Furthermore, S-TaLiRo provides a metric on how well a sample trace (system trajectory) satisfies the property, denominated ``robustness''. 
Hence, exploration methods looking for inputs that provide the best robustness value, can be applied in an automated manner, e.g., simulated annealing, cross-entropy, genetic algorithms and uniform random sampling. 

S-TaLiRo operates directly in MATLAB/Simulink, and numerous examples verifying performance requirements for complex control systems are provided. 

\subsection{SpaceEx}
SpaceEx specializes in analysis of hybrid systems with piecewise affine, non-deterministic dynamics, through computationally efficient reachability algorithms for the continuous transitions of the hybrid systems. 
In particular, SpaceEx computes ``flowpipe'' approximations of sets of reachable states. 
The hybrid system models to verify are constructed through a graphic interface, and saved in an xml format.

The restrictions on the continuous components of the hybrid systems, to piecewise affine dynamics, is shared by other tools for hybrid systems, such as the CheckMate model checker, and the theorem prover Keymaera~\cite{Chutinan,Jackson}. 
Consequently, we hypothesised that this tool, and other similar ones, would not be suitable to model and consequently verify the stability requirements of the discrete systems in Section~\ref{sc:examples}.

\section{EXPERIMENTS AND RESULTS}\label{sc:experiments}
For each system described in Section~\ref{sc:examples}, a model in the model checker's input language was developed and its verification attempted. 
All models are available online\footnote{\url{https://github.com/riveras/model_checking}} together with the results obtained. 
For all the linear systems, stable and unstable system parameters were applied to assess the correctness of the verification results. 

We evaluated the different model checking approaches according  to the following criteria: \textit{(a)} time to pre-process a model into a BDD representation, which is a critical aspect for symbolic model-checking; \textit{(b)} total checking time, i.e., time used to verify a requirement; \textit{(c)} amount of additional user input to specify parameters such as initial values, simulation time or loop iterations; and \textit{(d)} correctness of the translation semantics from Simulink to the input language of the model checker, if translation is necessary. 

\subsection{Experiments in NuSMV}
FSMs were developed manually, according to the semantics proposed in~\cite{Meenakshi}, for NuSMV version 2.4.1. 
On each FSM, the state variables transition sequentially according to each one of the operation blocks in the Simulink diagram,  starting from the delay. 
The state variables in the model were represented by 8 bits, to reduce the state space at the cost of inaccuracy and representation. 
We adjusted the basic operations to represent basic floating point numbers, also at great accuracy cost.

In the verification stage, initial parameters of $x=2$, $\mathbf{x}=[1;1]$ were applied for the scalar and matrix systems, respectively, signifying the verification of stability only for a single system's trajectory. 
For the first two scalar and matrix loops, parameters $a=\{0.9,1.9\}$ and $\mathbf{A}=\{[0.5 \quad 0; 0 \quad 0.5],[1.5 \quad 0; 0 \quad 1.5]\}$ were used for the stable and unstable versions. $K=[10.1,6]$ was employed for the unstable controlled system. 
No unstable version of the nonlinear system was verified. 
An LTL formula specified the stability requirement, over the Lyapunov function's difference, {\small \verb+G delay_lyap_output >= lyap_output+}. 

Results are shown in Table~\ref{tab:nusmv}, where YES denotes the compilation and the verification taking place, and NO denotes the failure to compile in less than two hours. T indicates the property is true, and F the property is false. Where the model compilation process succeeded, the size of the state space is indicated. 

\begin{table}[t]
\centering
\caption{Experiments in NuSMV}
\label{tab:nusmv}
\renewcommand{\arraystretch}{1.3}
\begin{tabular}{l|cc|c|cc|c}
\hline
\bf EXAMPLE & \multicolumn{3}{c|}{\bf STABLE SYSTEM} & \multicolumn{3}{c}{\bf UNSTABLE SYSTEM} \\
			& \multicolumn{2}{c|}{\bf Comp.?} & {\bf Verif.?}  & \multicolumn{2}{c|}{\bf Comp.?} & {\bf Verif.?} \\ \hline
Scalar loop	& 	YES	& $2^{120}$	&		YES: T			&  YES	&	$2^{120}$	&		YES: F	\\
Matrix loop	& 	YES	& $2^{160}$ 	&		YES: T			& 	NO	& --		& 		NO		\\
Controller	& 	NO	& -- 		&		NO				&   NO  	& --		&		NO	 	\\ 
Nonlinear	& 	YES	& $2^{176}$ 	&		YES: T			&   --	& --		&		--  		\\ \hline
\end{tabular}
\end{table}

\setcounter{paragraph}{0}
\paragraph{Time to pre-process a model into a BDD}
The results in Table~\ref{tab:nusmv} show the impressive size of the state-space, for relatively ``simple systems'', even with the reduced bit-vector size. 
NuSMV struggled to compute BDDs for systems that loop continuously without a fixpoint, such as the one of the unstable matrix linear system, within a reasonable time threshold. Consequently, verification cannot take place. 
This state-space explosion for unstable and the controlled loop is caused by the variables overflowing, as no related flags or added constraints were implemented to stop the infinite loops. 

\paragraph{Verification time}
The main overhead was caused by the building of the models into BDDs. 
When models were built, the verification time was less than one hour, although counterexamples took longer to compute when the property was found to be false.

\paragraph{Parameters to specify}
Ideally, as many  initial state  conditions as possible should be verified (i.e.\ ``all possible states''), representing different system trajectories. 
This process can be automated through scripting to control the models and NuSMV, although the chosen initial conditions will always be constrained by the bit-vector size. 
Nevertheless, this ``sampling'' of the initial conditions is not exhaustive over the state space. 

\paragraph{Correctness of the translation semantics}
The semantics in~\cite{Meenakshi} do not specify how to deal with floating-point operations and non-integer data, these are at the core of the semantics of Simulink. 
The provided arithmetic operations in NuSMV are not equivalent to the floating-point ones in Simulink. 
Furthermore, their semantics offer no guidance on how to correctly represent the functionality of a loop that increases continuously, as the bit-vectors would overflow without any implemented constraints. 

Overall, this approach is more suited to verify control systems at code implementation level, as proposed in~\cite{Herber2013}, providing adequate semantics (e.g., floating-point) and system loop constraints (to avoid overflows when variables reach their limits) are added, to emulate the system loop properties more closely within fixed width data types.
Nonetheless, the state-space sizes in the computation of BDDs are still an obstacle.

\subsection{Experiments in UCLID}
We manually adjusted the FSM models developed for NuSMV, according to UCLID's syntax. 
Following the semantics in~\cite{Herber2013}, a transition in the FSM is an unrolling of the whole system loop, going through all the serial block operations in the Simulink diagrams at once. 
The arithmetic division operation in the nonlinear system was approximated to 0, as the provided operators do not allow divisions with variable denominators as in NuSMV. 
We ran UCLID version 3.1, in a Fedora 6 Virtual Machine.

In the verification stage, the parameters for the stable and unstable systems were $a=\{0.9,12.9\}$ for the scalar loop, $\mathbf{A}=\{[0.5 \quad 0; 0 \quad 0.5],[13.5 \quad 0; 0 \quad 13.5]\}$ for the matrix loop, and $K=[11,6]$ for the unstable controlled system. 
The same initial values of NuSMV were used for the state variables, and a bit-vector size of 16 for all variables. 
Iteration bounds of $k=\{10,20,40,80\}$ were explored, i.e., the models were unrolled up to 80 times from the specified initial states. 
We verified the requirement through the computation of the expression {\small \verb+comparison := lyap.output <= delay2.output+} at each exploration step (``simulation'') of the model.

Table~\ref{tab:uclid} shows the results when checking the expression, for the different $k$ bounds. F indicates the expression was not true in at least one of the checks, and T indicates the expression was always true. 

\begin{table}[t]
\centering
\caption{Experiments in UCLID}
\label{tab:uclid}
\renewcommand{\arraystretch}{1.3}
\begin{tabular}{l|cccc|cccc}
\hline
\bf EXAMPLE 		& \multicolumn{4}{c|}{\bf STABLE SYSTEM} & \multicolumn{4}{c}{\bf UNSTABLE SYSTEM} \\ \hline
{\bf $k$ bound}	& {\bf 10} & {\bf 20} & {\bf 40}& {\bf 80} & {\bf 10} & {\bf 20} & {\bf 40}& {\bf 80}  \\ \hline
Scalar loop	& 	T			&	T	&  	T		&	T	&	F		&	F		&	F	&	F		\\
Matrix loop	& 	T			&	T	& 	T		& 	T	&	F		&	F		&	F	&	F		\\
Controller	& 	F			&	F	&   	F		&	F	&	F		&	F		&	F	&	F		\\ 
Nonlinear	& 	T			&	T	&	T		&	T	&	--		&	--		&	--	&	--		\\ \hline
\end{tabular}
\end{table}

\setcounter{paragraph}{1}

\paragraph{Verification time}
The verification process took seconds, as the models are unrolled iteratively according to the specified steps on-the-fly. 
Unfortunately, no counterexamples were provided when the expression checks failed.

\paragraph{Parameters to specify}
We specified initial values, as for NuSMV, and bounds on the number of exploration steps. 
It is not clear how to chose a suitable number of steps. However, this information is critical for verification since too small a number may lead to false positives if the expressions failed (i.e.\ are falsified) in the future.
Additionally, the expression checks at each execution step have to be encoded by hand explicitly, whereas in other model checkers this is done automatically by indicating a property to check.  

\paragraph{Correctness of the translation semantics}
As in NuSMV, the semantics in~\cite{Herber2013} do not consider floating-point operations, nor overflows. 
The impact of the lack of floating point built-in support is evident in the controlled system, where high precision multiplication operations are needed for an accurate computation of the requirement's expression value. 
Furthermore, the available arithmetic operations have limited functionality and there is no matrix data type, compared to NuSMV.

This approach is computationally less expensive than using NuSMV, allowing larger bit-vector sizes, although more limited in terms of arithmetic operations and data representation. 
Extending the operational functionalities and implementing some overflow constraints, this approach would be more suitable to the verification of control systems code, by exploring the loops for $k$ iterations, for both high-level functional and runtime requirements.
Pre- and post-conditions to avoid overflows and underflows would enable sound verification ``for all possible representable values'' of variables in a system loop, within bounded ranges and a fixed bit-vector widths.

\subsection{Experiments in S-TaLiRo}
For these experiments, we modified the Simulink diagrams presented in Section~\ref{sc:examples} by adding output probes to measure the Lyapunov function's difference over time, as required by the tool, S-TaLiRo version 1.61, running in MATLAB/Simulink version R2013a. 

For verification, ranges of $x=[-10 \quad 10]$ and $\mathbf{x}=[-10 \quad 10;-10 \quad 10]$ were specified for the initial state variable values, to be sampled by the tool. 
We used the same parameters for $a$ and $\mathbf{A}$ in the linear systems as in the NuSMV experiments, and an unstable controller of $-K$. 
The MTL properties to falsify were, for stable systems, ``the Lyapunov function's difference is eventually $>0$''; and, for unstable systems, ``the Lyapunov function's difference is always $<=0$''. 
We allowed 100 tests for different input samples, for three of the offered exploration methods to find traces that falsify the properties: simulated annealing (SA), cross-entropy (CE), and uniform random (UR). The cross-entropy method did not run successfully for the stable systems. 
The experiment configuration settings, initial or input value ranges, and MTL properties and exploration methods were specified via MATLAB scripts. 

The results are shown in Table~\ref{tab:staliro}, where T indicates a falsifying trace was not found (the system might not be unstable), and F indicates a falsifying trace was found (the system is unstable). 

\begin{table}[t]
\centering
\caption{Experiments in S-TaLiRo}
\label{tab:staliro}
\renewcommand{\arraystretch}{1.3}
\begin{tabular}{l|ccc|ccc}
\hline
\bf EXAMPLE & \multicolumn{3}{c|}{\bf STABLE SYSTEM} & \multicolumn{3}{c}{\bf UNSTABLE SYSTEM} \\\hline
{\bf Sampling}	& {\bf SA} & {\bf CE}  & {\bf UR} & {\bf SA} & {\bf CE} & {\bf UR}  \\ \hline
Scalar loop	& 		T		&	--		&  	T	&		F 	&	F		&	F		\\
Matrix loop	& 		T		&	--		&  	T	&		F 	&	F		&	F		\\
Controller	& 		T		&	--		&  	T	&		F 	&	F		&	F		\\
Nonlinear	& 		T		&	--		&	T	&		--	&	--		&	--		\\ \hline
\end{tabular}
\end{table}

\setcounter{paragraph}{1}
\paragraph{Verification time} 
The verification process took less than a minute in total, for the specified number of samples and diagram simulation time per sample. 
This time is expected to increase if more samples and larger value ranges are introduced. 

\paragraph{Parameters to specify} 
S-TaLiRo requires specifying either ranges for the initial values of the state variables, or ranges for the inputs (if any), the number of samples, the simulation time per run (sample), and the exploration method, to mention some of the most important parameters. 
Some of these parameters can be intuitively tuned, but others depend more on understanding of the tool's functionality and previous knowledge.

\paragraph{Correctness of the translation semantics} 
The ``translation'' process from our Simulink diagrams involves adding input and output block probes, to indicate which are the input signals to sample, and which are the output signals that the property to verify refers to. 

Overall, this approach is the most straightforward to use over our Simulink diagrams, since no translation is needed, and the amount of additional specifiable parameters is reduced. 
Additionally, S-TaLiRo allowed to cover more of the possible initial state variable values in an automated manner, compared to having to change the initial values manually in the models used in UCLID and NuSMV. 
Nevertheless, this approach is not complete for the verification of stability in general, since it does not offer a proof in the case a system is stable -- i.e., the results for the MTL property of ``eventually unstable'' for a stable system only indicate that the system is not unstable within the sampled initial state values and simulated time. 
Alternatively, it could be the case that the requirement of ``always stable'' is not falsified in unstable systems, if the initial state values and simulation time interval are of a trajectory that appears to converge to an equilibrium point. 

Statistical model checking through S-TaLiRo allows the verification of all types of systems (e.g., continuous, discrete, linear, nonlinear, stochastic and delayed) as long as they are modelled in Simulink. 
Although, this approach cannot substitute the strength of computing a proof of a requirement ``for all possible variable values and initial states'', from theorem proving (e.g., in~\cite{Araiza:2015}), it is suitable as a method to search for evidence that a requirement is not satisfied. 

\subsection{Experiments in SpaceEx}

We attempted to construct models for the systems in Section~\ref{sc:examples}, but the syntax did not allow expressing difference equations and nonlinear terms, including the Lyapunov functions. 
In contrast, continuous systems similar to the discrete ones, such as
\begin{equation}
\dot{x}=-0.5x,\nonumber
\end{equation}
were easily constructed. 
Stability can only be verified, for continuous or hybrid systems, through computing the reachability of the models, given initial state conditions, $x(0)$, specified as single values or intervals. If the system is stable, the system's reachability is bounded by these initial conditions, and the system converges (reaches a fixpoint) to the equilibrium point in $x=0$. If the system is unstable, the system's state variables diverge and no fixpoint is reached. 
The same concepts extend to multi-variable linear continuous systems. 

SpaceEx has the potential to be extended to include discrete systems and their respective ``flowpipe'' algorithms, once developed. 

\section{CONCLUSIONS AND FUTURE WORK}\label{sc:conclusion}

We explored the verification of stability for discrete systems' designs in Simulink, through different model checking variants and corresponding state-of-the-art tools. 
We aimed to find how well these different model checking variants are suited to this particular problem domain, how compatible they are with respect to Simulink diagrams as a system modelling language, what are their limitations in practice, and if any of them would substitute or complement our previously developed theorem proving approach~\cite{Araiza:2015}, by providing examples as evidence of requirement violation. 

We explored four model checking variants through related tools, symbolic, bounded, statistical and hybrid, verifying stability based on Lyapunov methods for discrete linear and nonlinear systems. 
Our experiments and results provide an insight on the strengths and limitations of these model checking approaches with respect to the verification of stability of discrete systems. 

We found that statistical model checking through S-TaLiRo is the most suitable option to complement our previous work on automatic theorem proving.
This same approach is the most compatible with Simulink diagrams of systems described as difference equations, compared to symbolic model checking with NuSMV or bounded model checking with UCLID -- based on available translation semantics from Simulink to NuSMV and UCLID--, and hybrid model checking with SpaceEx. 

In the future, we will incorporate model checking into our automatic theorem proving methodology~\cite{Araiza:2015}. This will allow us to return evidence as proof (in the form of counterexamples) when the systems are not stable or do not satisfy other control or performance requirements, thus enhancing the usability of the approach and facilitating debug. 
Our evaluation shows that S-TaLiRo is the ideal candidate for this extension.

Instead of developing individual tools to verify the same requirement over different types of systems, software platforms could be extended to recognize the system's characteristics and apply relevant algorithmic variants, according to sound theoretical frameworks.
MATLAB/Simulink could be used as platform to encode standardized models (i.e., through the same graphical language), connected to theorem provers, external constraint, satisfiability and optimization solvers. 

The computation of Lyapunov functions for all kinds of systems remains a research challenge. 
A plausible alternative is to further explore model checking related computational techniques, such as reachability approximations or BMC combined with SMT solvers as in~\cite{Zoucav}, to find ``stable'' regions of the state space~\cite{Duggirala}, or regions that satisfy other performance requirements. 
Other alternatives point towards the use of statistical model checking or sampling and search methods as in S-TaLiRo, combined with optimization problems, to compute Lypaunov functions and barrier certificates~\cite{Kapinski} for larger sets of types of systems.  
We will be exploring such alternatives in the future. 

\section*{Acknowledgements}
The work presented in this paper was supported by the EPSRC grant EP/J01205X/1 RIVERAS: Robust Integrated Verification of Autonomous Systems.

\bibliographystyle{IEEEtran}

\bibliography{references}

\end{document}